\documentclass[conference, 10pt, letter]{IEEEtran}
\IEEEoverridecommandlockouts
\usepackage{cite}
\usepackage{amsmath,amssymb,amsfonts, bm, multirow}
\usepackage{algorithmic}
\usepackage{graphicx}
\usepackage{textcomp}
\usepackage{xcolor}
\usepackage{epsfig, bm, mathdots}

\def\BibTeX{{\rm B\kern-.05em{\sc i\kern-.025em b}\kern-.08em
    T\kern-.1667em\lower.7ex\hbox{E}\kern-.125emX}}
\usepackage[hidelinks]{hyperref}

\usepackage{color}
\usepackage{graphicx}
\usepackage{latexsym}
\usepackage{enumerate}
\usepackage{caption}
\usepackage{subcaption, hyperref, setspace, tikzsymbols, textcomp, dsfont}

\DeclareSymbolFont{extraup}{U}{zavm}{m}{n}

\usepackage{amsthm, algorithm, algorithmic}

\newcounter{pb}
\newcounter{spb}
\newfam \msbfam
\font \tenmsb=msbm10    \textfont \msbfam=\tenmsb

\newcommand{\C}{\mathbb C}

\newcommand{\E}{\mathbb E}

\def\jj{{\mathfrak{j}}}

\def\va{{\bm{a}}}
\def\vb{{\bm{b}}}

\def\vh{{\bm{h}}}

\def\vq{{\bm{q}}}
\def\vr{{\bm{r}}}
\def\vs{{\bm{s}}}

\def\vw{{\bm{w}}}
\def\vx{{\bm{x}}}

\def\mB{{\bm{B}}}
\def\mC{{\bm{C}}}

\def\mF{{\bm{F}}}

\def\mH{{\bm{H}}}
\def\mI{{\bm{I}}}

\def\mS{{\bm{S}}}

\def\mV{{\bm{V}}}
\def\mW{{\bm{W}}}
\def\mX{{\bm{X}}}
\def\mY{{\bm{Y}}}

\theoremstyle{definition}

\newtheorem{Fact}{Fact}
\newtheorem{assum}{Assumption}

\usepackage{titlesec}
\titlespacing*{\section}{0pt}{1em}{0em}
\titlespacing*{\subsection}{0pt}{0em}{0em}
\titlespacing*{\subsubsection}{0em}{0em}{0em}
\def\va{{\bm a}}

\def\vs{{\bm s}}
\def\vx{{\bm x}}

\def\vh{{\bm h}}
\def\vq{{\bm q}}

\def\mH{{\bm H}}

\def\mY{{\bm Y}}

\def\jj{{\mathfrak{j}}}

\def\va{{\bm{a}}}
\def\vb{{\bm{b}}}

\def\vh{{\bm{h}}}

\def\vq{{\bm{q}}}
\def\vr{{\bm{r}}}
\def\vs{{\bm{s}}}

\def\vw{{\bm{w}}}
\def\vx{{\bm{x}}}

\def\mB{{\bm{B}}}
\def\mC{{\bm{C}}}

\def\mF{{\bm{F}}}

\def\mH{{\bm{H}}}
\def\mI{{\bm{I}}}

\def\mS{{\bm{S}}}

\def\mV{{\bm{V}}}
\def\mW{{\bm{W}}}
\def\mX{{\bm{X}}}
\def\mY{{\bm{Y}}}

\def\jj{{\mathfrak j}}

\allowdisplaybreaks

\titlespacing*{\section}{0pt}{3pt}{0pt}
\titlespacing*{\subsection}{0pt}{3pt}{0pt}
\definecolor{orange}{RGB}{255,107,0}

\begin{document}
\title{
One-Bit Sigma-Delta DFRC Waveform Design:\\
Using Quantization Noise for Radar Probing
\vspace{-0pt}
}

\author{
\IEEEauthorblockN{
Wai-Yiu Keung$^{\dagger \mathsection}$, 
Hei Victor Cheng$^{\ast}$, 
and 
Wing-Kin Ma$^{\mathsection}$ 
 \vspace{10pt}
}
\IEEEauthorblockA{\small
${}^{\dagger}$Department of Computer Science and Engineering, The Chinese University of Hong Kong, Hong Kong SAR of China\\ 
${}^{\mathsection}$Department of Electronic Engineering, The Chinese University of Hong Kong, Hong Kong SAR of China \\
${}^{\ast}$Department of Electrical and Computer Engineering, Aarhus University, Denmark
}
 \vspace{-15pt}
}

\maketitle
\begin{abstract}
Dual-functional radar-communication (DFRC) signal design 
has received much attention lately. 
We consider the scenario of one-bit massive  multi-input multi-output (MIMO)
wherein one-bit DACs are employed for the sake of saving hardware costs. 
Specifically, a spatial Sigma-Delta \((\Sigma\Delta)\) modulation scheme 
is proposed for one-bit MIMO-DFRC waveform design. 
Unlike the existing approaches which require
large-scale binary optimization,
the proposed scheme performs \(\Sigma\Delta\) modulation on 
a continuous-valued DFRC signal.
The subsequent waveform design is formulated as a constrained least square problem, 
which can be efficiently solved. 
Moreover, 
we leverage quantization noise for radar probing purposes, 
rather than treating it as unwanted noise.
Numerical results demonstrate that
the proposed scheme performs well in
both radar probing and downlink precoding.
\end{abstract}

\section{Introduction}
Integrated sensing and communication (ISAC) 
has recently received significant interests from academia and industry\cite{liufan2018jointTX, zhang2021overview,  liu2022isac}. 
In the past, 
radar and communication were treated as
independent subjects,
and they usually operate in different frequencies. 
With the increasing demand for spectral resources,
it is anticipated that 
the next-generation communication system 
will support spectrum sharing for several functions 
in the same frequency band. 

Sharing bandwidth for radar and communication systems 
is not new in civil and commercial applications.
For example, the air traffic control radar probing signal operates in the {\it IEEE L-band},
while the LTE standard occupies the same band for mobile communications \cite{wang2017spectrum}.
Such a kind of spectrum sharing is usually achieved by means of interference management,
which requires some prior exchange between the two (radar and communication) systems \cite{zheng2019rccoexist}.
What distinguishes ISAC from the coexisting approach
is the ability to use a single signal for both radar and communication purposes.
Such an integrated approach 
reduces the information overhead between the two systems.
It also reduces the operating cost for the overall system as 
it only requires one set of hardware equipment to implement both systems.
This has motivated the study of 
DFRC waveform design, 
particularly under the multiple-input multiple-output (MIMO) settings
\cite{liufan2018optWaveform, liufan2022crb}. 

On the other hand,
the subject of quantized signal processing 
has attracted attention in communication, 
especially in massive MIMO precoding \cite{li2018massive, sohrabi2018onebit, shao2019framework}.
This is motivated by the fact that 
using high-resolution analog-to-digital/digital-to-analog converters (ADCs/DACs) in large MIMO systems is too expensive, and they consume too much energy. 
In the context of precoding, the majority of the existing literature seek
to design the transmitted signal by optimizing a certain quality-of-service metric,
subject to the one-bit signaling constraint.
There are some pioneering works on one-bit optimization-based MIMO-DFRC signal design \cite{cheng2021onebitDFRC, yan2022onebitDFRC, yu2022onebitDFRC, wu2024onebitDFRC}. 
Such schemes were numerically found to yield good performance.
However, the main challenge is that large-scale discrete optimization is required, 
and that incurs significant computational overheads.

In addition to the discrete optimization-based approach,
there is a line of studies that analyze the effect of one-bit quantization 
over the transmitted MIMO signal \cite{li2017onebitDAC, saxena2017onebitDAC}.
As direct quantization incurs large distortion,
the question is whether we can suitably control the quantization noise 
to enhance precoding performance. 
More recently, 
spatial $\Sigma\Delta$ modulation 
has been applied to coarsely quantized  
massive MIMO precoding \cite{shao2019onebit, keung2024cvx}.
Temporal $\Sigma\Delta$ modulation is a classic 
ADC/DAC structure that features quantization noise-shaping  \cite{aziz1996sigdel}.
Assuming that the base station (BS) uses a uniform linear array (ULA),
spatial $\Sigma\Delta$ modulation shapes the quantization noise away from the broadside.
Therefore, when the BS sends a $\Sigma\Delta$ one-bit signal, 
the users experience little distortion if they are in
the small angle range, e.g., $[-20^\circ, 20^\circ]$.
The merit of $\Sigma\Delta$ precoding is that
it allows the use of traditional simple precoding designs (e.g., zero-forcing) 
that require far less computation power than the discrete optimization designs. 
Apart from precoding, 
spatial $\Sigma\Delta$ modulation has found other applications, such as,
MIMO detection, channel estimation, power amplifier distortion mitigation, and phase design for reconfigurable intelligent surface 
\cite{nguyen2024sigdelMIMODet, sankar-chepuri-2022, liu2023spatial, keung2023transmitting}.

{
In this paper, we consider spatial $\Sigma\Delta$ modulation for one-bit MIMO-DFRC waveform design,
wherein one-bit DACs are employed at the DFRC base station. 
Specifically, 
we perform quantization on a pre-designed
DFRC signal by
the first-order $\Sigma\Delta$ modulator.
Our pre-modulated DFRC signal design is
formulated as a constrained least square problem, 
which can be efficiently solved. 
We will show that
the proposed one-bit $\Sigma\Delta$ DFRC waveform design
achieves good performances in both
radar probing and precoding.}


\section{Problem Statement}
\label{sec:problem-statement}

\begin{figure}[t!] 
\centering
\centerline{\includegraphics[width=.9\linewidth]{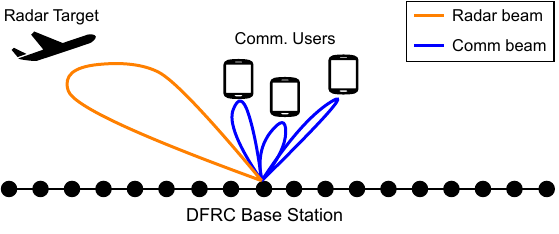}}
\vspace{-0pt}
\caption{System Diagram}
\label{fig:setting}
\end{figure}

Consider a MIMO-DFRC base station with $N$ transmit antenna 
arranged as a uniform linear array (ULA). 
We aim to design signals
to serve $K$ single-antenna downlink communication users.
Meanwhile, there is a single point target, 
which we want to probe for radar detection purposes.
The scenario is illustrated in Fig.~\ref{fig:setting}.
Let $\vx_\ell$
be the transmitted DFRC signal at the $\ell$-th time instant.
The signal $\mX = [\vx_1, \dots, \vx_L]\in \C^{N \times L}$
is designed to serve both radar and communication purposes.

\subsection{MIMO Radar Probing}
Let us first describe the MIMO radar probing task.
Under the narrow-band flat-fading channel assumption, 
the transmitted pattern at angle $\theta \in (-90^\circ, 90^\circ)$ is
\[
    r_\ell(\theta) = \va_\theta ^H \vx_\ell
\]
with $\va_\theta = (1, e^{ \jj \omega}, \dots, e^{  \jj \omega (N-1)}) \in \C^N, \omega =\frac{ 2\pi d}{\lambda} \sin(\theta),$
being the steering vector.
Here, $d \leq \lambda/2$ is the inter-antenna spacing, and 
$\lambda$ is the wavelength of the carrier.
The emitted signal pattern over the $L$ instants can be written as 
\[
    \vr^\top_\theta = \va_\theta^H\mX_{\rm R}  \in \C^L
\]
where $\mX_{\rm R} = [\vx_1, \dotsm \vx_L] \in \C^{N \times L}$
is the transmitted probing signal matrix.
The transmitted beampattern is given by
\begin{equation}
    \label{eqn:spatial-beampattern}
     P(\theta) 
            = \mathbb{E} \left(|\va_\theta^H \vx_\ell|^2\right) 
            = \va_\theta^H \mC \va_\theta
\end{equation}
where $\mC = \E[\vx_\ell \vx_\ell^H]$ is the covariance matrix of the transmitted signal, 
which we aim to design. 

Assume that the radar target 
falls within a region of interest $\Theta_{M} \subset (-90^\circ, 90^\circ)$.
Denote $\theta_0$ as the midpoint of $\Theta_M$.
We call $\Theta_M$ as the mainlobe.
Let $\Theta_S \subset  (-90^\circ, 90^\circ) \setminus \Theta_M$ 
be a pre-defined sidelobe.
Given a per-antenna transmission power budget $p$, 
one can design the beampattern by, e.g.,
maximizing the worst-case mainlobe-to-sidelobe ratio,
while controlling the mainlobe ripple. 
This is formulated as \cite{stoica2007probing}:
\begin{equation}\label{eqn:tx-beampatt-opt}
    \begin{array}{r l l}
         \underset{\mC \succeq 0, \tau \geq 0}{\rm max}& \; \tau\\
         {\rm s.t. } & P(\theta_0)-P(\theta) \geq \tau, & \theta \in \Theta_{S}\\
                    & (1+\epsilon) P(\theta_0) \geq P(\theta) \geq (1-\epsilon) P(\theta_0), & \theta \in \Theta_{M}\\
                    & [\mC]_{nn} \leq p, \quad n = 1,\dots, N.
    \end{array}
\end{equation}
Here, $\epsilon \geq 0$ is a small constant that controls the mainlobe ripple.
As can be seen, problem \eqref{eqn:tx-beampatt-opt}
aims to maximize the radiation power gap
between $\Theta_M$ and $\Theta_S$. 
Thus, the strength of the probing signal 
in $\Theta_S$ will be smaller than that in $\Theta_M$.
The idea is to avoid clutters in $\Theta_S$ from 
reflecting strong echo that may confuse (or interfere)
the radar detector. 
It is worth noting that problem \eqref{eqn:tx-beampatt-opt} is a convex optimization problem.

Given a properly designed covariance matrix $\mC$,
there are several ways to generate $\mX_{\rm R}$,
see, e.g., \cite{stoica2007probing,stoica2008wavefomr}. 
One simple way is to synthesize 
\begin{equation}
\label{eqn:gaussian-randomizer}
    \mX_{\rm R} = \mC^{1/2} \mW
\end{equation}
where $\mC^{1/2}$ is the positive semidefinite square root of the matrix $\mC$;
$\mW = [\vw_1, \dots, \vw_L]$,
and $\vw_\ell \sim \mathcal{CN}({\bm 0}, \frac{1}{\sqrt{2}}\mI_N)$
is a standard complex Gaussian random vector.
We should mention that eqn. \eqref{eqn:gaussian-randomizer}
has little control on the signal values in $\mX_{\rm R}$. 
In practice,
it requires more dedicated algorithms to search for a $\mX_{\rm R}$
that satisfies the one-bit waveform constraint.

\subsection{MIMO Communication Precoding}
Next, we describe the downlink transmission task.
Let $\mX_{\rm C} \in \C^{N\times L}$ be the precoded signal matrix to be designed.
Let the downlink symbol stream for the $k$-th user be $\vs_k = (s_{k, 1}, \dots, s_{k, L})^\top$.
The symbol matrix for all the $K$ users within the channel coherence block
can be written as $\mS = (\vs_1, \dots, \vs_K) \in \mathcal{S}^{K\times L}$,
where $\mathcal{S}$ is a constellation set, e.g., the $M$-ary QAM set.
The received signal model is given by 
\[
    \mY = \mH \mX_{\rm C} + {\mV}_{\rm C}
\]
where $\mH = [\vh_1, \dots, \vh_K]^H \in \C^{K \times N}$ is the channel matrix;
$\mV_{\rm C} \in \C^{K \times L}$ with
its entries being $v_{k, \ell}\sim \mathcal{CN}(0, \sigma_v^2)$ is the background noise.  
A simple precoding design is to minimize the multi-user interference (MUI), 
while making sure that the signal matrix satisfies some transmission constraints,
e.g., the one-bit signal constraint.
This can be formulated as 
\begin{equation}
    \label{eqn:one-bit-MUI-minimization}
    \begin{array}{r l l}
         \underset{\mX_{\rm C}}{\rm min}& \;  \|\mH\mX_{\rm C} - \mS\|_F^2 \\
         {\rm s.t.}& \; \mX_{\rm C} \in \{\pm1, \pm\jj\}^{N\times L}
    \end{array}.
\end{equation}   
When there is no constraint on $\mX_{\rm C}$, 
one can simply use the zero-forcing scheme $\mX_{\rm C} = \mH^\dagger \mS$
to perform precoding. 
Under the one-bit constraint, the problem is a large-scale binary optimization problem, which is challenging to solve. 

\subsection{One-Bit MIMO-DFRC Signal Design}
We have reviewed the designs of the MIMO radar probing signal matrix $\mX_{\rm R}$
and the MIMO communication precoding matrix $\mX_{\rm C}$.
We now turn to the design of $\mX$ 
that works for both radar probing and communication precoding. 
Let $\delta \in (0, 1)$ be a pre-defined trade-off factor between the radar and communication purposes. 
Suppose that an ideal radar probing matrix $\mX_{\rm R}$ is synthesized by, e.g.,
\eqref{eqn:tx-beampatt-opt}--\eqref{eqn:gaussian-randomizer}.
Following  \cite{liufan2018optWaveform}, we pose the one-bit DFRC signal design problem as:
\begin{equation}
    \label{eqn:DFRC-formulation}
    \begin{array}{r l l}
         \underset{\mX}{\rm min}& \; \delta \, \|\mH\mX - \mS\|_F^2 + (1-\delta)\|\mX - \mX_{\rm R}\|_F^2\\
         {\rm s.t.}& \; \mX \in \{\pm1, \pm\jj\}^{N\times L}
    \end{array}.
\end{equation}   
Such optimization seeks to design a binary matrix $\mX$
that balances between MUI reduction and proximity to the 
desired radar matrix $\mX_{\rm R}$.
Define
\begin{equation}
\label{eqn:def-matrix-FB}
     \mF = \left[
    \begin{array}{c}
         \sqrt{\delta}\mH \\
         \sqrt{1-\delta}\mI_N  
    \end{array}
    \right]
    \text{ and }
    \mB = \left[
    \begin{array}{c}
         \sqrt{\delta}\mS\\
         \sqrt{1-\delta}\mX_{\rm R}  
    \end{array}
    \right],
\end{equation}
the objective function of \eqref{eqn:DFRC-formulation}
can be expressed as
\begin{equation}
    \label{eqn:DFRC-LS-form}
        \begin{array}{r l l}
         \underset{\mX}{\rm min}& \; \|\mF \mX - \mB \|_F^2\\
         {\rm s.t.}& \; \mX \in \{\pm1, \pm\jj\}^{N\times L}
    \end{array}.
\end{equation}
This problem is challenging due to the one-bit constraint. 

As before, if the binary constraint is relaxed, 
the problem admits
a closed-form unconstrained least-square solution of $\mX^{\star} = \mF^\dagger \mB$.
However, such a solution lies in the free-space $\C^{N\times  L}$. 
We can obtain a one-bit signal matrix by 
putting $\mX^{\star}$ through a direct quantizer, 
i.e., taking directly the sign of the real and imaginary part of $\mX^{\star}$.
The drawback is that the quantization error may be large, 
which may cause significant performance degradation in both radar and communication.


\section{Spatial Sigma-Delta Modulation}
\label{sec:sigdel}

\begin{figure}[t!] 
\centering
\centerline{\includegraphics[width=.87\linewidth]{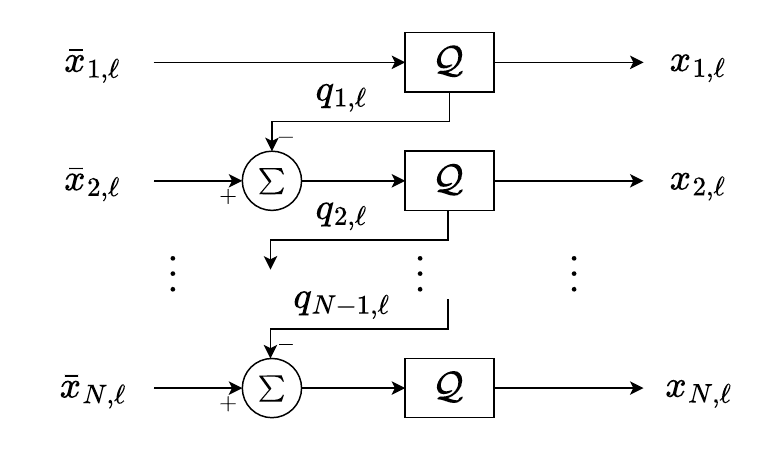}}\vspace{-0pt}
\caption{Spatial $\Sigma\Delta$ modulator.}
\label{fig:sigdel-modulator}
\end{figure}

In this paper, 
we are less interested in the discrete optimization approaches
for MIMO DFRC waveform design, as their high computation complexity prohibits them to be implemented in practice. Instead, we emphasize on a precode-then-quantize scheme.
In particular, we propose to use spatial $\Sigma\Delta$ modulation
to shape the quantization noise power away from the communication user, 
and to leverage it for radar-probing purposes in the region of interest. 
To explain such spatial noise-shaping technique in one-bit MIMO,
we describe the idea of spatial $\Sigma\Delta$ modulation from a communication point of view. 

A spatial $\Sigma\Delta$ modulator is depicted in Figure~\ref{fig:sigdel-modulator}.
The modulator input $\{\bar{x}_{n, \ell}\}$ is a pre-defined free-space signal. 
The complex quantizer is defined by 
\[
    \mathcal{Q}(x) = {\rm sgn}(\Re(x)) + \jj {\rm sgn}(\Im(x)),
\]
which quantizes the in-phase and quadrature components of the input signal. 
As seen from the system diagram, 
the quantization noise is passed to the adjacent antenna, 
forming a spatial feedback loop.
The system can be written as
\begin{align}\label{eqn:sigdel-sys}
    \vx_\ell &= \mathcal{Q}(\bar{\vx}_\ell - \vq_\ell^-)
           = \bar{\vx}_\ell - \vq_\ell^- + \vq_\ell
\end{align}
where 
\(\vq_\ell = (q_{1, \ell}, \dots,q_{N, \ell})\) 
is the quantization noise sequence; and
\(\vq_\ell^- = (0, q_{1, \ell}, \dots,q_{N-1, \ell})\) 
is the delayed quantization noise at the feedback path.
Before we proceed, 
we should also mention the stability criterion of the system \eqref{eqn:sigdel-sys}.
Notice that the quantization noise depends on the feedback noise:
\begin{equation}
\label{eqn:noise-dependence}
        q_{n, \ell} = x_{n, \ell} - (\bar{x}_{n, \ell} - q_{n-1, \ell})
\end{equation}
for $n = 1, \dots, N$. 
The amplitude of $|q_{n, \ell}|$ may grow large with $n$. 
This phenomenon may happen if the quantizer in the modulator is overloaded.
To prevent that, 
it is typical to limit the input signal amplitude by
\[
    -1 \leq {\Re(\bar{x}_{n, \ell})}, {\Im(\bar{x}_{n, \ell})} \leq 1.
\]
One can show that this results in $q_{n, \ell} \in [-1, 1]\times[-\jj, \jj]$. 
We refer readers to \cite{aziz1996sigdel, shao2019onebit, keung2024cvx} for detail. 
This no-overload condition has a strict amplitude restriction on the input signal,
which can be achieved by signal normalization in the digital domain for DAC applications.
As a heuristic, we further describe the following fact:\\

\begin{Fact}{\bf (Probably no-overload condition.)}
    Suppose that $ \bar{x}_{n, \ell}$'s are 
    complex Gaussian random variables following $\mathcal{CN}(0, \sigma^2_{\bar{x}})$ for $\ell = 1, \dots, L$. 
    If $\sigma_{\bar{x}}^2 \leq 2/9$, 
    then the $\Sigma\Delta$ modulator has no overload for $99.7\%$ of the samples. \\
\end{Fact}
\vspace{-3mm}
\noindent The idea of this fact follows from the famous three-sigma rule \cite[pp. 378]{proakis1996dsp}.
In essence, this allows us to perform signal design on $\{\bar{x}_{n, \ell}\}$
by constraining its spatial variance instead of using a strict amplitude bound.

We now discuss the core insight of spatial noise-shaping.
Consider a downlink communication user locating at angle $\theta \in (-90^\circ, 90^\circ)$
with unit channel gain.
The noiseless received signal model is given by
\begin{equation}\label{eqn:noise-shaping}
    \va_\theta^H \vx_\ell 
        \simeq \va_\theta ^H \bar{\vx}_\ell
          + (1-e^{-\jj \omega})\left(  \sum_{n = 1}^{N-1} q_{n-1, \ell}  e^{-\jj \omega (n-1)}\right),
\end{equation}
which approximately holds when $N \gg 1$.
Here, $\omega = \frac{2\pi d}{\lambda}\sin(\theta)$ is the spatial frequency associated with the user angle. 
We observe that the quantization noise perceived at the angle $\theta$
is shaped by a high-pass filter $|1-e^{-\jj \omega}|$. 
This implies that 
the power of quantization noise 
in a small angular region (e.g., $\theta \in [-20^\circ, 20^\circ]$) will be small,
and hence users facing the array broadside will be less affected by the quantization effect. 
We also see that a smaller $d$ will help to reduce the quantization noise perceived by the user.
Nevertheless, a very small $d$ is prohibitive as it may induce mutual coupling. 
It is common to use $\lambda/8 \leq d\leq \lambda/2$ in the current literature.  

Prior works
have shown that spatial $\Sigma\Delta$ modulation
can be used to generate quantized MIMO signals to serve communication users in a small angular range \cite{shao2019onebit, keung2024cvx}. 
It is also suggested that the quantization noise power at high angular region
is rich. 
This motivates us to make use of the shaped quantization noise
for radar-probing purposes at large angles. 

\section{Proposed Scheme}
\label{sec:proposed-scheme}
In this section, we present the proposed design for
$\Sigma\Delta$ DFRC signals. 
Our proposed scheme is separated into two steps. 
First, we provide a covariance-based probing signal design strategy 
customized for the pre-$\Sigma\Delta$ modulation signal matrix.  
Then, we generate signals from the designed covariance matrix
that serves as an ideal radar probing signal $\bar{\mX}_{\rm R}$. 
Second, we will design a pre-$\Sigma\Delta$ modulation 
signal for DFRC purposes. 
We will make use of the 
synthesized pre-modulated radar signal $\bar{\mX}_{\rm R}$ to
perform DFRC signal design by a variant of \eqref{eqn:DFRC-formulation}.
Finally, 
the designed DFRC signal will be converted to one-bit signals by
spatial $\Sigma\Delta$ modulation.

\subsection{Covariance-Based Probing Signal Design}
\label{sec:sigdel-probing-scheme}
In this subsection, we reformulate the 
transmit beampattern design problem \eqref{eqn:tx-beampatt-opt}
by taking the effect of spatial $\Sigma\Delta$ modulation into consideration.
If the one-bit transmitted signal is generated by \eqref{eqn:sigdel-sys}, 
the spatial beampattern is expressed as
\begin{equation}
    \label{eqn:sigdel-beampatt}
    P(\theta) = \mathbb{E}(|\va_\theta^H(\bar{\vx}_\ell +\vq_\ell - \vq_\ell^-)|^2).
\end{equation}
Due to the dependence of $\vq_{\ell}$ on $\bar{\vx}_\ell$ (cf. eqn. \eqref{eqn:noise-dependence}),
quantization noise analysis for $\Sigma\Delta$ modulation is a difficult task.
A convenient approximation, which is widely adopted in the $\Sigma\Delta$ literature, 
is stated as follows:

\begin{assum}
    The quantization noise $q_{n, \ell}$'s are
    i.i.d. random variables, and they are
    uniformly distributed on the support
    $[-1, 1]\times [-\jj, \jj]$.
\end{assum}

Under such assumption,
we simplify \eqref{eqn:sigdel-beampatt} to
\begin{equation*}
    P(\theta) = \mathbb{E}(|\va_\theta^H \bar{\vx}_\ell |^2) 
                +D(\theta),
\end{equation*}
where $D(\theta) = \mathbb{E}(|\va_\theta^H(\vq_\ell - \vq_\ell^-)|^2)$.
Observe that the equation structure is identical to \eqref{eqn:noise-shaping}.
Therefore, we can write 
\begin{align*}
    D(\theta)
    &\simeq 
    \textstyle 
    |1-e^{-\jj \omega}|^2 
    \mathbb{E}\left(\left| \sum_{n = 1}^{N-1} q_{n-1, \ell}  e^{-\jj \omega (n-1)}\right|^2\right)\\
    &= \textstyle
    \frac{8(N-1)}{3}\left|\sin \middle(\frac{\pi d \sin(\theta)}{\lambda}\middle)\right|^2  
\end{align*}
wherein we have used $\mathbb{E}(|q_{n, \ell}|^2) = 2/3$,
which is a consequence of Assumption 1. 
The spatial beampattern of the one-bit $\Sigma\Delta$ modulated signal is written as 
\begin{equation}
    \label{eqn:sigdel-beampatt-final-expression}
    P(\theta) = \va_\theta^H \bar{\mC} \va_\theta + D(\theta)
\end{equation}
where $\bar{\mC}$ is the correlation matrix of the pre-$\Sigma\Delta$ signal $\bar{\vx}_\ell$. 
Plugging this into \eqref{eqn:tx-beampatt-opt},
we modify the beampattern design problem 
to fit for
the design of $\bar{\mC}$ as: 
\begin{equation}\label{eqn:tx-beampatt-sigdel}
    \begin{array}{r l l}
         {\rm max}& \; \tau\\
         {\rm s.t. } &\va_{\theta_0}^H \bar{\mC} \va_{\theta_0}\!-\!\va_{\theta}^H \bar{\mC} \va_{\theta}\!+\!D(\theta_0)\!-\!D(\theta)  \geq \tau, & \theta \in \Theta_{S}\\
         &(1\!+\!\epsilon)  (\va_{\theta_0}^H  \bar{\mC} \va_{\theta_0}\!+\!D(\theta_0))
            \!\geq\!\va_{\theta}^H   \bar{\mC} \va_{\theta}\!+\!D(\theta) 
                     \\
         & \hspace{20pt}\geq (1\!-\!\epsilon)  (\va_{\theta_0}^H  \bar{\mC} \va_{\theta_0}\!+\!D(\theta_0))
                &\theta\in\Theta_{M}\\
         & [\bar{\mC}]_{n,n} \leq p, \quad n = 1,\dots, N\\
         &\bar{\mC} \succeq 0, \tau \geq 0
    \end{array}
\end{equation}
Similar to problem \eqref{eqn:tx-beampatt-opt},
the customized problem \eqref{eqn:tx-beampatt-sigdel}
is a convex problem,
and can be handled by an off-the-shelf solver, such as {\tt cvx} \cite{cvx}. 
Note that the covariance $\bar{\mC}$ can be pre-computed and stored for, 
e.g., several probing directions,
in practice.
Once $\bar{\mC}$ is determined, 
we can generate the ideal radar-probing signal matrix $\bar{\mX}_{\rm R}$ by
using \eqref{eqn:gaussian-randomizer}.


\subsection{DFRC Waveform Design}
\label{sec:dfrc-sigdel-scheme}
We now describe the design of the pre-$\Sigma\Delta$ modulated 
DFRC signal matrix.
Given a communication-to-radar trade-off factor $\delta$,
a user symbol matrix $\mS$, and
a desired probing matrix $\bar{\mX}_{\rm R}$,
we consider the DFRC design for pre-$\Sigma\Delta$ modulated signal:
\begin{equation}
    \label{eqn:DFRC-sigdel-constraint}
        \begin{array}{r l l}
         \underset{\bar{\mX}\in \C^{N\times L}}{\rm min}& \; \|\mF \bar{\mX} - \mB \|_F^2\\
         {\rm s.t.} & \; {\rm diag}(\bar{\mX}^H\bar{\mX}) \leq \frac{2N}{9} {\bm{1}}
    \end{array}
\end{equation}
whereas $\mF, \mB$ are as defined in \eqref{eqn:def-matrix-FB}, with $\mX_{\rm R}$ replaced by $\bar{\mX}_{\rm R}$;
\(\bar{\mX} =  [\bar{\vx}_1, \dots, \bar{\vx}_L]\) is the pre-$\Sigma\Delta$ modulated signal matrix; 
and the constraint is to ensure that the spatial signals $\bar{\vx}_\ell$'s 
have spatial variances smaller than $2/9$ by Fact 1. 
{ 
    Here, we assumed $\bar{\vx}_\ell$'s are
    independent zero-mean complex Gaussian random vectors
    implicitly.
}
Problem \eqref{eqn:DFRC-sigdel-constraint} is a convex problem
with a large problem size ($N$ is large in massive MIMO).
In practice, the optimization is conducted in each channel coherence block.
Thus, it is undesirable to solve problem \eqref{eqn:DFRC-sigdel-constraint} using a general purpose solver.

We are therefore interested in analyzing the problem structure of \eqref{eqn:DFRC-sigdel-constraint}.
We first observe from \eqref{eqn:DFRC-sigdel-constraint} 
that the variables can be decomposed into $L$ vectors, 
and rewrite
\begin{equation}
    \label{eqn:DFRC-sigdel-constraint-per-ell}
        \begin{array}{r l l}
         \underset{\{\bar{\vx}_\ell \in \C^N\}_{\ell = 1}^{L} }{\rm min}& \; \sum_{\ell = 1}^{L} \|\mF \bar{\vx}_\ell - \vb_\ell\|_2^2 \\
         {\rm s.t.} & \; \|\bar{\vx}_\ell\|_2^2 \leq 2N/9, 
         \quad \ell = 1, \dots, L,
    \end{array}
\end{equation}
where $\vb_\ell$ is the $\ell$-th column of $\mB$.
From here, we see that each of the constraints concerns only one $\bar{\vx}_\ell$,
and that the minimum of each $\|\mF\bar{\vx}_\ell - \vb_\ell \|_F^2$ shall 
lead to the minimum of the objective in \eqref{eqn:DFRC-sigdel-constraint}. 
Thus, we design $\bar{\mX}$ in a column-by-column fashion.
Specifically, for each $\ell = 1, \dots, L$, we solve 
\begin{equation}
    \label{eqn:DFRC-sigdel-constraint-per-ell-2}
        \begin{array}{r l l}
         \underset{\bar{\vx}_\ell \in \C^N }{\rm min}& \; \|\mF \bar{\vx}_\ell - \vb_\ell\|_2^2 \\
         {\rm s.t.} & \; \|\bar{\vx}_\ell\|_2^2 \leq 2N/9
    \end{array}.
\end{equation}
Problem \eqref{eqn:DFRC-sigdel-constraint-per-ell-2} is a $2$-norm constrained least square problem.
It is well-known that the optimal solution to \eqref{eqn:DFRC-sigdel-constraint-per-ell-2} admits the form
\[
    {\bar{\vx}}_\ell^\star = (\mF^H\mF + \bar{\lambda}_\ell \mI)^{-1} \mF^H \vb_\ell  
\]
where $\bar{\lambda}_\ell$
can be efficiently solved by, e.g., using singular value decomposition and bisection search \cite[Ch. 6.2.1]{golub2013matrix}.

We should also remark on some operation aspects in 
using $\Sigma\Delta$ modulation for one-bit DFRC signal design.
First, we describe the communication aspect. 
Same as the preceding study \cite[Sec. 5.4]{shao2019onebit}, 
we assume the channel vectors $\vh_k$'s follow the
multi-path channel model
\begin{equation}
    \label{eqn:multi-path-channel}
    \vh_k = \textstyle \sum_{j = 1}^{J} \alpha_{k, j} \va_{\theta_{k, j}}
\end{equation}
where $J$ is the number of multi-paths for each user;
$\alpha_{k, j}$ and $\theta_{k, j}$ are, respectively, 
the complex channel gain and the angle for the $j$-th path of the $k$-th user.
We assume the path angles $\theta_{k, j}$'s are relatively small,
so that the prescribed spatial noise-shaping works well for the communication users
\cite[Sec. 4.2]{liu2023spatial}. 
Second, 
we assume the region of interest (mainlobe of the beampattern)
for radar probing is within a higher angular region, e.g., 
$|\theta|\geq 50^\circ$. 
This is because the higher angle has richer noise power as shaped by the $\Sigma\Delta$ modulator,
which can naturally be used for probing purposes.
{
In addition, it has been shown in \cite{shao2019onebit, keung2023transmitting, keung2024cvx} that 
the error feedback filter in the $\Sigma\Delta$ modulator can be 
customized to serve communication users 
beyond the broadside, such as $[20^\circ, 60^\circ]$.
Nevertheless, we will use the above-mentioned basic first-order $\Sigma\Delta$ modulator
for the paper's coherency. 
}

\section{Numerical Results}
This section provides numerical results to 
evaluate the performance of the proposed scheme.
Some settings are described as follows.
The BS is a ULA with $N = 256$ antennas,
and the inter-antenna spacing is set as $d = \lambda/8$.
In all test cases, 
the transmitted signal has
amplitude $\leq 1$ in both real and imaginary parts. 
For the beampattern synthesis aspects, 
the angular domain $[-90^\circ, 90^\circ]$ is 
uniformly sampled with $0.5^\circ$ step size. 
The mainlobe is set as $\Theta_M = [55^\circ, 65^\circ]$,
which means that the center angle is $\theta_0 = 60^\circ$. 
The mainlobe ripple tolerance is $\epsilon = 0.1$. 
The sidelobe is set as $\Theta_S = [-90^\circ, 50^\circ]\cup[70^\circ, 90^\circ]$.
For communication aspects,
the number of downlink users is $K = 6$,
and the multi-path number is $J = 4$.
The symbol stream is drawn from the $4$-QAM constellation set.
The user angles $\theta_{k, j}$'s 
are uniformly drawn from the sector $[-20^\circ, 20^\circ]$,
and they are separated from each other by at least $0.5^\circ$. 
The phase of $\alpha_{k, j}$ is randomly drawn from $[-\pi, \pi]$,
and its magnitude is generated by 
$|\alpha_{k, j}| = r_0/r_1$, where $r_0 = 10$ and $ r_1 \in [20, 100]$.
The block length is set to $L = 300$.
The reported results were obtained by Monte-Carlo simulations
with $1000$ trials.

\begin{figure*}[t!]
\centering
\begin{minipage}{.31\textwidth}
  \begin{subfigure}{\linewidth}
    \centering
    \includegraphics[width=.99\linewidth]{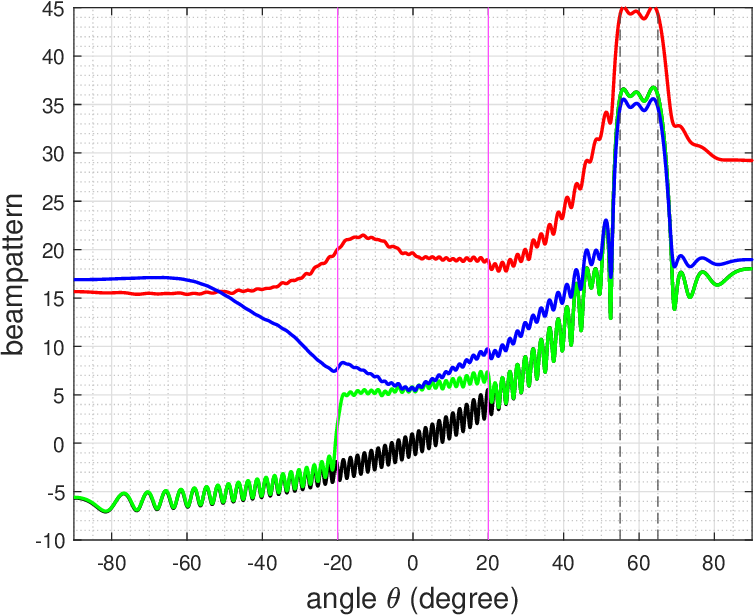}
    \caption{Transmitted Beampattern, $\delta = 0.1$.}
    \label{fig:beampatt-9C1R}
  \end{subfigure}\\   
  \vspace{10pt}
  
  \begin{subfigure}{\linewidth}
    \centering
    \includegraphics[width=.99\linewidth]{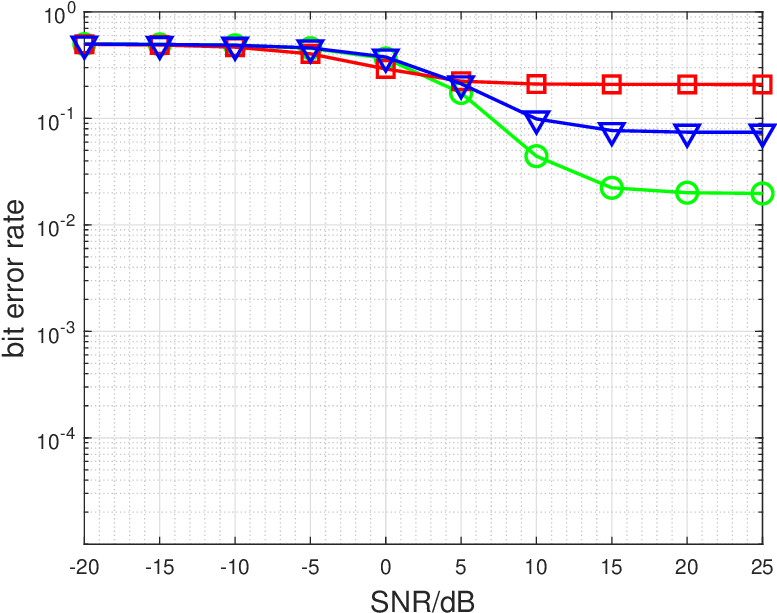}
    \caption{Bit Error Rate, $\delta = 0.1$.}
    \label{fig:ber-9C1R}
  \end{subfigure}
\end{minipage}%
\hfill 
\begin{minipage}{.31\textwidth}
  \begin{subfigure}{\linewidth}
    \centering
    \includegraphics[width=.99\linewidth]{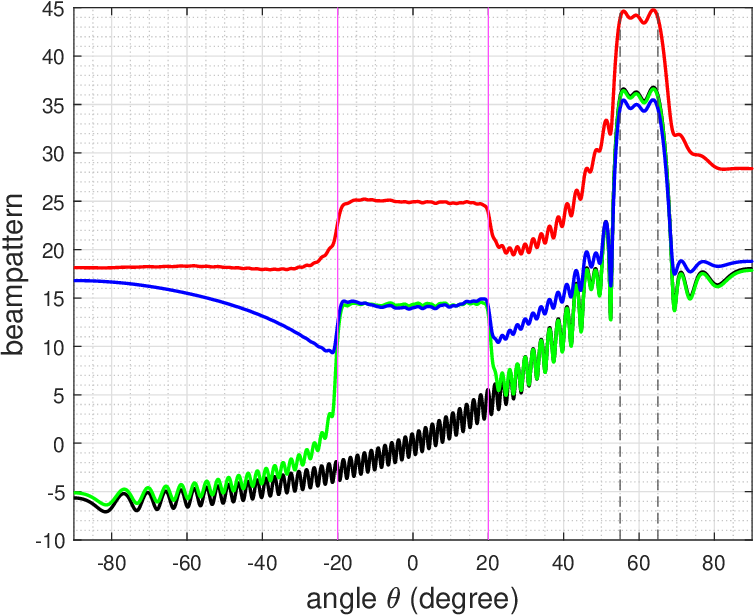}
    \caption{Transmitted Beampattern, $\delta = 0.5$.}
    \label{fig:beampatt-5C5R}
  \end{subfigure}\\   
  \vspace{10pt}
  
  \begin{subfigure}{\linewidth}
    \centering
    \includegraphics[width=.99\linewidth]{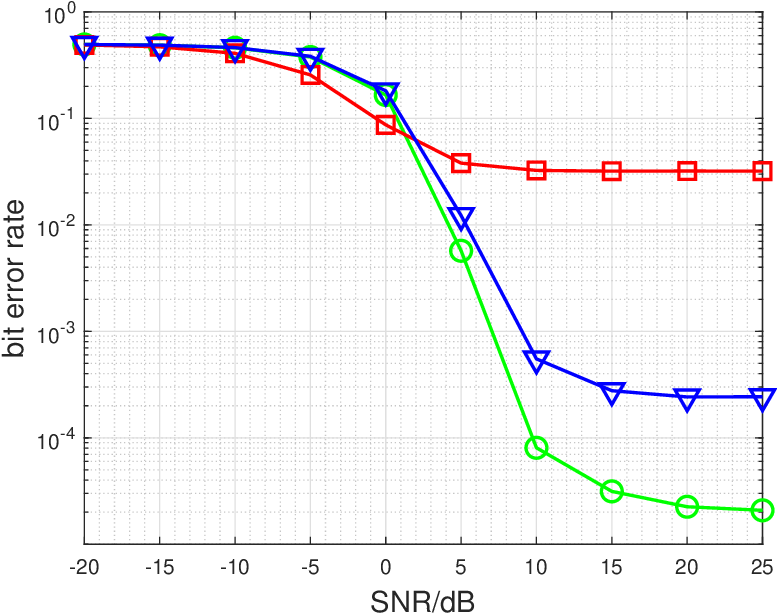}
    \caption{Bit Error Rate, $\delta = 0.5$.}
    \label{fig:ber-5C5R}
  \end{subfigure}
\end{minipage}
\hfill 
\begin{minipage}{.31\textwidth}
  \begin{subfigure}{\linewidth}
    \centering
    \includegraphics[width=.99\linewidth]{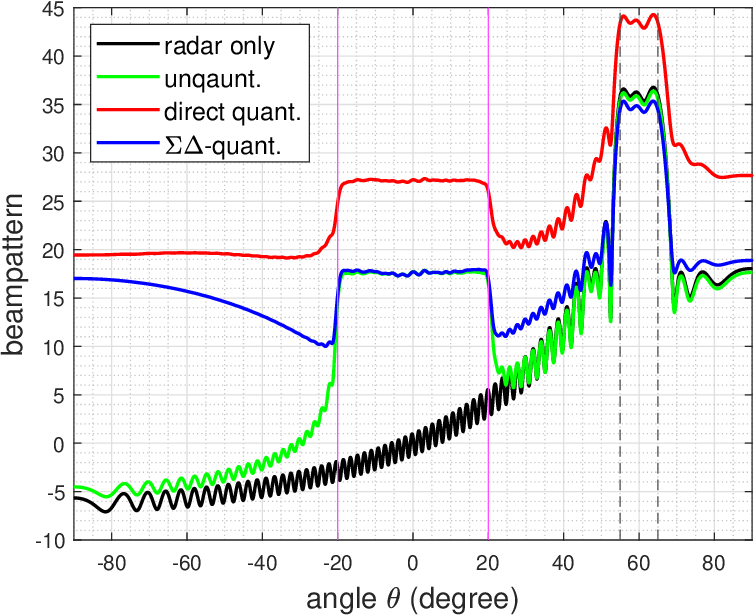}
    \caption{Transmitted Beampattern, $\delta = 0.9$.}
    \label{fig:beampatt-1C9R}
  \end{subfigure}
  \\   
  \vspace{10pt}

  \begin{subfigure}{\linewidth}
    \centering
    \includegraphics[width=.99\linewidth]{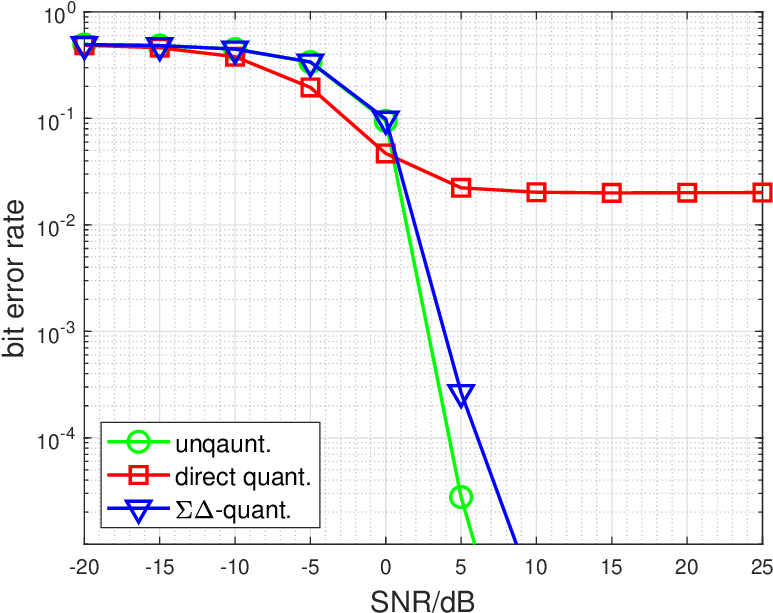}
    \caption{Bit Error Rate, $\delta = 0.9$.}
    \label{fig:ber-1C9R}
  \end{subfigure}
\end{minipage}
\\     \vspace{20pt}
  
\caption{Performances of the proposed DFRC signal design scheme. 
``radar only": The unquantized radar only signal $\mX_{\rm R}$,
``unquant.": The unquantized benchmark, 
``direct quant.": The direct 1-bit quantization of unquant., 
``$\Sigma\Delta$-quant.": The proposed $\Sigma\Delta$ modulation scheme. 
}
\label{fig:sim-sigdel-dfrc}
\end{figure*}

We test the proposed $\Sigma\Delta$-DFRC scheme
with different communication-to-radar trade-off factors $\delta$.
Figure~\ref{fig:sim-sigdel-dfrc} shows the beampattern and the bit error rate curves of the proposed scheme.
We also benchmark our scheme with the unquantized DFRC waveform design 
in \cite{xu2022experimental}
and with its direct one-bit quantized counterpart. 
The results for $\delta = 0.1$, $\delta = 0.5$ and $\delta = 0.9$
are captured in Figs.
(\ref{fig:beampatt-9C1R})--(\ref{fig:ber-9C1R}), 
(\ref{fig:beampatt-5C5R})--(\ref{fig:ber-5C5R}),
and (\ref{fig:beampatt-1C9R})--(\ref{fig:ber-1C9R}),
respectively.

Consider the transmitted beampatterns in Figs.~(\ref{fig:beampatt-9C1R}), (\ref{fig:beampatt-5C5R}), and (\ref{fig:beampatt-1C9R}).
We use magenta lines to indicate the user angle region, 
and black dashed lines to indicate the mainlobe.
We see that all the beampatterns have 
a strong spike over the mainlobe $[55^\circ, 65^\circ]$. 
We observe that when $\delta$ is large, 
there is an increasing energy level within the communication user angle region. 
We also see that the proposed $\Sigma\Delta$ scheme 
is closer to the radar-only waveform than the direct quantization scheme.
To give the reader more quantitative results, 
Table~\ref{tab:RBP_MSE} shows the mean squared error (MSE in dB) between 
the DFRC beampattern and the radar-only beampattern. 
We see that the proposed $\Sigma\Delta$ scheme gives a smaller MSE 
than direct quantization in all settings. 

Next, we evaluate the precoding performance by examining the bit error rates.
The results are illustrated in Figs.~(\ref{fig:ber-9C1R}), (\ref{fig:ber-5C5R}), and (\ref{fig:ber-1C9R}).
In all test cases,
we see that the unquantized scheme 
performs the best in the high SNR region;
and that the direct quantized scheme 
failed to suppress the bit error rate curve. 
Also, the proposed $\Sigma\Delta$ scheme works as expected.
Despite being a one-bit signal,
it loses only a few dB from the unquantized benchmark,
except for the error flooring effect in the high SNR region.

\begin{table}[t]
    \begin{center}
        \begin{tabular}{r|c c c}
         &                  $\delta = 0.1$ & $\delta = 0.5$ &   $\delta = 0.9$ \\\hline
        unquant. &               $12.8889$&     $32.8019$&      $39.5047$\\
       direct quant. &           $75.9298$&      $74.9555$&     $73.7323$\\
       $\Sigma\Delta$ quant. &   $47.5657$  &  $48.1901$ &   $48.9220$
    \end{tabular}
    \end{center}
    \caption{MSE (in dB) between the DFRC beampatterns and the radar-only beampattern.}
    \label{tab:RBP_MSE}
\end{table}

\section{Conclusions}
This paper presents a 
$\Sigma\Delta$ modulation scheme for one-bit DFRC signal design. 
Under some operational conditions, 
we leverage the $\Sigma\Delta$ noise-shaping property
to turn the unwanted quantization noise  
into part of the useful radar probing signal. 
The pre-$\Sigma\Delta$ modulated signal matrix design
is formulated as a constrained least square problem,
which can be efficiently solved.
Our numerical result shows that the 
proposed one-bit $\Sigma\Delta$-DFRC scheme 
performs well in both radar probing
and MIMO precoding.


\bibliographystyle{IEEEtran}
\bibliography{ref}

\end{document}